\documentclass{iauc}
\usepackage{graphicx}
\usepackage{natbib}

\title[ESP with AMBER] %% give here short title %% 
{Extrasolar Planets with AMBER/VLTI,
 What can we expect from current performances ?}

\author[Millour, Petrov et al.] %% give here short author list %%
{
  F.~Millour$^{1;2}$,
  M.~Vannier$^{4}$, 
  R.~G.~Petrov$^{1}$,
  B.~Lopez$^{3}$ \and
  F.~Rantakyr\"o$^{4}$
}

\affiliation{$^1$ Laboratoire Universitaire d'Astrophysique de Nice - U.M.R.
  6525\break
  Universit\'e de Nice-Sophia Antipolis, Parc Valrose,
  06108 Nice Cedex 02, France \\[\affilskip]
  $^2$Laboratoire d'Astrophysique de Grenoble, U.M.R. 5571\break
  Universit\'e Joseph Fourier/C.N.R.S., BP 53, F-38041 Grenoble Cedex
  9, France\\[\affilskip]
  $^3$Laboratoire Gemini, U.M.R. 6203\break
  Observatoire de la C\^ote d'Azur/C.N.R.S.,
  Avenue Copernic, 06130 Grasse, France \\[\affilskip]
  $^4$European Southern Observatory\break
  Casilla 19001, Santiago 19, Chile
}

\pubyear{2005}
\volume{xxx} %% insert here IAU Symposium No.
\pagerange{000--000}
\date{?? and in revised form ??}
\jname{Proceedings Title IAU Colloquium }
\editors{A.C. Editor, B.D. Editor and C.E. Editor, eds.}
\begin{document}

\maketitle

\begin{abstract}
  We present the current performances of the AMBER / VLTI
  instrument in terms of differential observables (differential phase
  and differential visibility) and show that we are already able to
  reach a sufficient precision for very low mass companions spectroscopy
  and mass characterization. We perform some extrapolations with the
  knowledge of the current limitations of the instrument facility.

  We show that with the current setup of the AMBER instrument, we
  can already reach $3\sigma = 10^{-3}$ radians and have the
  potential to some low mass companions characterization (Brown
  dwarves or hypothetical very hot Extra Solar Giant
  Planets). With some upgrades of the VLTI infrastructure,
  improvements of the instrument calibration and improvements of the
  observing strategy, we will be able to reach $3\sigma = 10^{-4}$
  radians and will have the potential to perform Extra Solar Giant
  Planets spectroscopy and mass characterization.

  % There are still limitations of the infrastructure
  % that
  % will be probably solved in a moderately short amount of time (about 1
  % to 2 years)

  %% add here a maximum of 10 keywords, to be taken form the file <Keywords.txt>.
  \keywords{techniques: interferometric, stars: planetary systems,
    instrumentation: high angular resolution}

\end{abstract}

\firstsection % if your document starts with a section,
\section{Introduction}

In this paper we discuss the current highest performances of
Colour-Differential Interferometry (CDI) on the AMBER instrument and
compare these performances to signal amplitude we computed from
low-mass companion simulations.

CDI is based on {\itshape simultaneous} interferometric observations
in different spectral channels. As a high-angular resolution and
high-dynamic technique, it presents two major advantages. First, the
chromatic differences in visibility and phases are much less sensitive
to instrumental and atmospheric instabilities, and therefore are
easier to calibrate than the absolute complex visibility. Since the
beginning of long-baseline optical interferometry with separated
apertures, many early astrophysical results have been obtained using
this self-calibration feature
\citep{1986A&A...165L..13T,1989Natur.342..520M}. Second, the
colour-differential phase can be measured with an accuracy much better
than the angular interferometric resolution $\lambda/B$. For objects
much smaller than the diffraction limit, it is proportional to the
variation of the object photocentre with wavelength.

This paper is placed in the context of Extra Solar Planets
characterization with interferometry \citep{2005-ESP-MNRAS} and is
intending to show that this technique has already the potential to get
some scientific results on high contrast binaries.

\section{Differential Observables computation and error bars}

\subsection{The differential phase estimator}

The expression of the coherent flux on tha AMBER instrument is \citep{2004SPIE.5491.1222M}:

\begin{equation}
  % \begin{multline}
  C(t,\lambda) = 2N(\lambda) V_i(t,\lambda) V(\lambda) \sqrt{p_1(t,\lambda) p_2(t,\lambda)} \sqrt{\sum_{k=1}^{N_x} a_{1k}(\lambda) a_{2k}(\lambda)} \times\\
  e ^{\phi_{i}(t,\lambda) + \phi_{p}(t,\lambda) + \phi_{o}(t,\lambda) + \phi_{c}(t,\lambda)} \label{C}
  % \end{multline}
\end{equation}

In this equation $k$ is the pixel index (spatial direction),
$N(\lambda) $ is the unknown object's flux, $p_1(t,\lambda) $ and
$p_2(t,\lambda) $ are transmission coefficients for the two combined
beams, and $a_{1k}(\lambda)$ and $a_{2k}(\lambda)$ are related to
specific features of each pixel (shape of the beam).  $V_i(t,\lambda)$
is the instrumental contrast and $V(\lambda) $ is the amplitude of the
complex visibility. 

$\phi_{p}(t,\lambda)$ is the phase induced by the achromatic piston,
$\phi_{i}(t,\lambda)$ is the instrumental-induced phase that varies
with time (since the fixed part is already removed by the data
reduction algorithm of AMBER as explained in
\citet{2004SPIE.5491.1222M}), $\phi_{o}(t,\lambda)$ is the observed
object's phase and $\phi_{c}(t,\lambda)$ is the chromatic phase
induced by other causes (Chromatic atmospheric phase for example).

% The expression of the phase $\phi_{p}(t,\lambda)$ is $\phi_{p}(t,\lambda) = \frac{2 \pi
%   \delta(t) }{ \lambda}$ with $\delta(t)$ the achromatic piston.

If we suppose that $\phi_{i}(t,\lambda) = 0$ (no variable instrumental
phase), $\phi_{c}(t,\lambda) = 0$ (no chromatic effect) and
$\phi_{o}(t,\lambda) = 0$ (unresolved or centro-symmetric object) then we
can correct the complex coherent flux from the achromatic piston effect
by:

\begin{equation}
  C_{\rm nop}(t,\lambda) = C(t,\lambda) \times e^{\frac{- 2 i \pi \delta(t) }{ \lambda}} \label{CpstCorr}
\end{equation}

One can note that we need an estimation of the achromatic piston for
each sample of time. This is a part of the Ph.D. thesis of Eric Tatulli
\citep{2004-These-Tatulli} and it will not be explained in detail in this
article.
We compute a reference channel for each spectral channel, taking care
of non biasing the interspectral term by removing the ``work''
spectral channel before averaging:

% When we cannot compute the achromatic piston (low RSB data),
% a way to get the differential phase is to suppose that the
% achromatic piston term is very small in regard to the coherence length
% of the fringe pattern or to use different techniques that will not be
% described in this article.

\begin{equation}
  C_{\rm ref}(t,\lambda_{k}) = \left< C_{\rm nop}(t,\lambda_i) \right>_{\lambda_i \neq \lambda_{k}} \label{Cref}
\end{equation}

We compute then the interspectral term between this reference spectral channel and
the work spectral channel:

\begin{equation}
  W(\lambda_{k}) = \left< \frac{C_{\rm nop}(t,\lambda_{k})^{}C_{\rm ref}(t,\lambda_{k})^*}
    {|C_{\rm ref}(t,\lambda_{k})|^2} \right>_{(t)} \label{Interspectrum}
\end{equation}

Then we compute the differential phase:

\begin{equation}
  \phi_{\rm diff}(\lambda) = \arg (W(\lambda)) \label{phiDiff_resPist}
\end{equation}

This expression of the differential phase is quite accurate with a
good achromatic piston correction. That is why it can be
applied only to high flux sources (for example the star 51 Peg. has a
K magnitude of 5, which is sufficient for this type of application). We
can express the interspectral term by:

\begin{equation}
  W(\lambda) = |W(\lambda)| e ^{i \phi_{\rm diff}(\lambda)} \label{Intersp_autre_form}
\end{equation}

Please note that when $\phi_{o}(t,\lambda)$ cannot be neglected
with regards to $\phi_{p}(t,\lambda)$, then the expression of the
differential phase is different and contains a bias related to the
interferometric phase, which leads to a bias in $W(\lambda) $. For
well resolved objects, this effect has to be taken into account and
leads to a specific treatment. Thus it is well beyond the scope of
this paper and will be discussed in a futher one. 

\subsection{The differential visibility estimator}

Going back to the interspectral term evaluation \ref{Interspectrum},
we see that its modulus can be expressed as: $\frac{V(\lambda)}
{V_{\rm ref}}$. If we have an unbiased estimate of this modulus (which
we call $\widetilde{V_{\rm diff}}(\lambda)$), then we can compute the
differential visibility. This estimate can be made with the real part
of the differential phase correction of the interspectral term:

\begin{equation}
  \widetilde{V_{\rm diff}}(\lambda) = \Re (W(\lambda) \times e ^{ - i 
    \phi_{\rm diff}(\lambda) })
\end{equation}

This estimate of the differential visibility is unbiased as we expect
the visibility and phase to be uncorrelated.
% (see figure \ref{complexUnbiasedEstimator}).

% \begin{center}
%   \begin{figure}[htpb]
%     \begin{center}
%       \includegraphics[width=0.4\hsize, angle=-90]{complexNr.pdf.ps}
%       \caption[]{Example showing why the real part of the complex
%         interspectral term is an unbiased estimator.}
%       \label{complexUnbiasedEstimator}
%     \end{center}
%   \end{figure}
% \end{center}

\subsection{The error bars}

\subsubsection{In theory}

Starting from the theoretical estimation of the differential phase, we
can express the differential phase and visibility noises from the
fundamental photon $\sqrt{N_*}$, thermal $\sqrt{N_{\rm th}}$ and
detector $\sigma_{\rm RON} \sqrt{n_{\rm f} \, n_{\rm pix}}$ noises
\citep{1989dli..conf..249P}.

\begin{equation}
  \sigma_{\Phi} = \frac{\sqrt{ (N_* + N_{\rm th} + n_{\rm f}\,n_{\rm pix}\,\sigma_{\rm RON}^2)/2 }}{V\, \langle N \rangle }
  \label{noise_phi}
\end{equation}

\begin{equation}
  \sigma_{V}= \frac{\sqrt{ N_* + N_{\rm th} + n_{\rm f}\,n_{\rm pix}\,\sigma_{\rm RON}^2 }}{\langle N \rangle }
  \label{noise_V}
\end{equation}

For information, the closure phase noise is given by:

\begin{equation}
  \sigma_{\psi}= \sqrt{3} \times \sigma_{\Phi}
\end{equation}

\subsubsection{in practice}

For the practical error computation, we perform a statistical
dispersion of the observed points and assume a gaussian noise. The
expression of the estimated noise for each observable is then the
standard deviation of the measurements over an exposure:

\begin{equation}
  \sigma_{X}= \sqrt{ \frac{ \sum _ {1} ^ {n_{\rm f}} (X - \langle X
      \rangle)^2 }{n_{\rm f}^2}}
  \label{statNoise}
\end{equation}

\section{Typical expected signal}\label{sec:esp}

\subsection{Simulating the companion}

We simulated a standard binary star model as in eq. \ref{eqnStar} and
applied the computation we currently use on the AMBER instrument to
extract differential visibilities, differential phases and closure
phase, computed the error bars assuming an average instrumental
contrast of 50\%, 1000 frames of integration, a detector
noise of 11$e^-$ per pixel and per frame and a total photon
count of $4.2 \times 10^7$, the same figures as in the following
section.

\begin{equation}
  C_{jk}(\lambda) = \frac{1 + R(\lambda) e^{-2i \pi \overrightarrow{u_{jk}} \cdot \overrightarrow{\rho} } }{1 + R(\lambda) }
  \label{eqnStar}
\end{equation}

We used a modeled exoplanet spectrum from \citet{2001ApJ...556..885B}
convolved with the resolution of the AMBER instrument in LR mode ($R
\approx 35$). We scaled this flux ratio to a maximum value of
$\approx 10^{-3}$ in order to estimate the signal amplitude of a 10
times brighter companion.

% \begin{figure}[htbp]
%   \centering
%   \begin{tabular}{c}
%     \includegraphics[width=0.5\textwidth, angle=0]{SIM_RapFlux1.ps}
%   \end{tabular}
%   \caption{
%     \footnotesize{
%       Simulated flux ratio between the star and the low mass companion
%       where we used a simulated spectrum of \citet{2001ApJ...556..885B}
%       scaled at a maximum flux ratio of 1e-3 in order to simulate a 10
%       times brighter companion.
%     }
%   }
%   \label{figFluxRat}
% \end{figure}

\begin{figure}[htbp]
  \centering
  \begin{tabular}{ccc}
    \includegraphics[width=0.31\textwidth, angle=0]{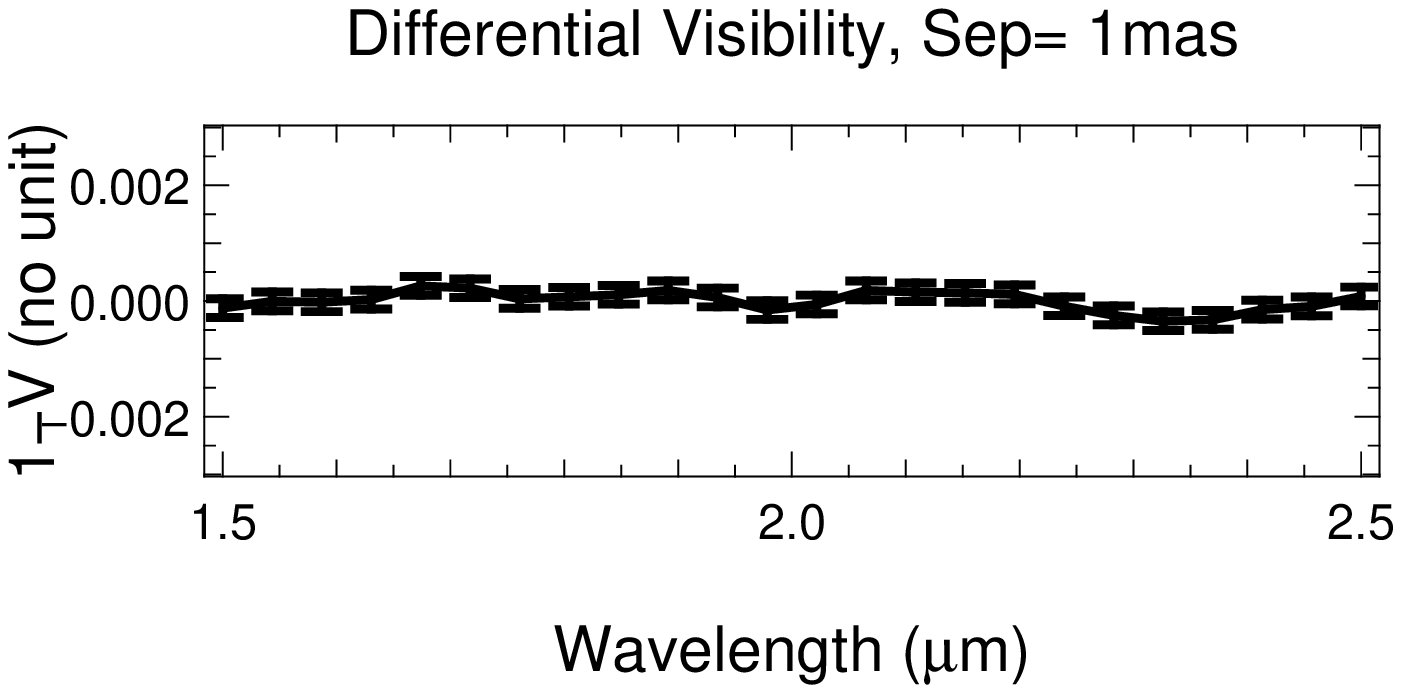}&
    \includegraphics[width=0.31\textwidth, angle=0]{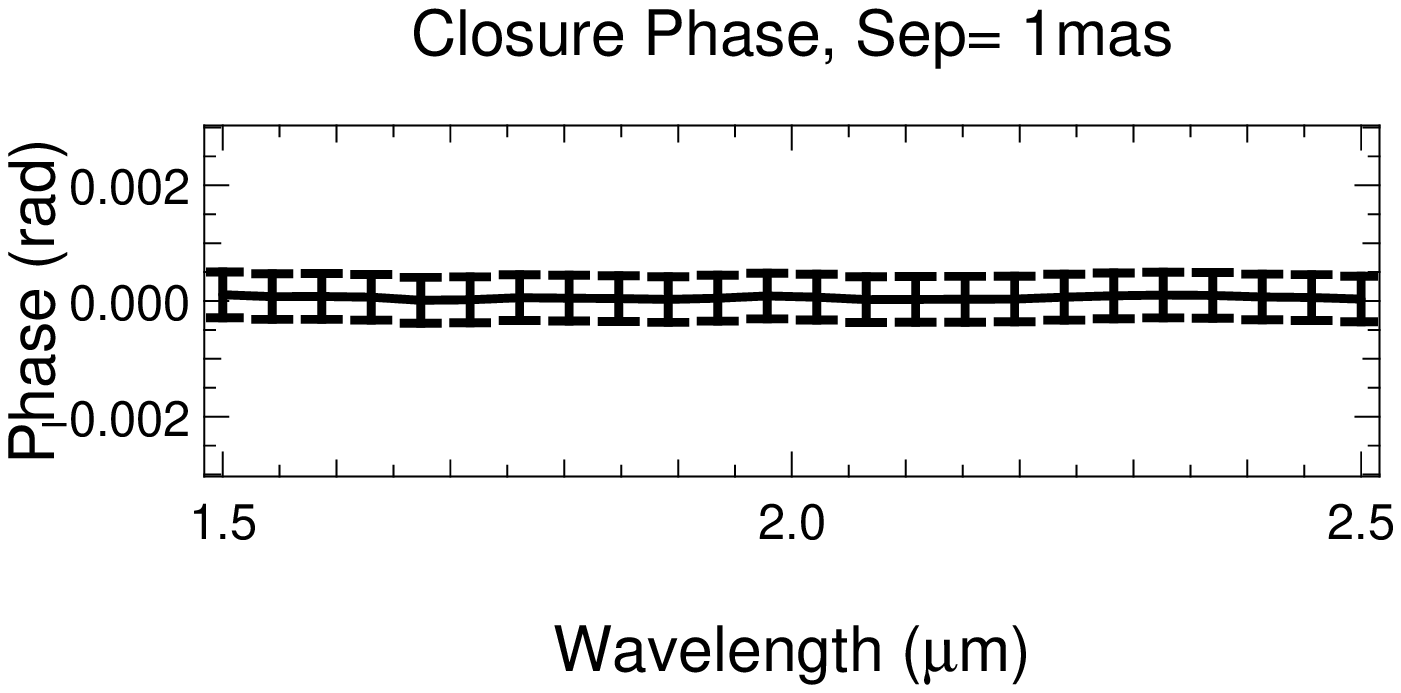}&
    \includegraphics[width=0.31\textwidth, angle=0]{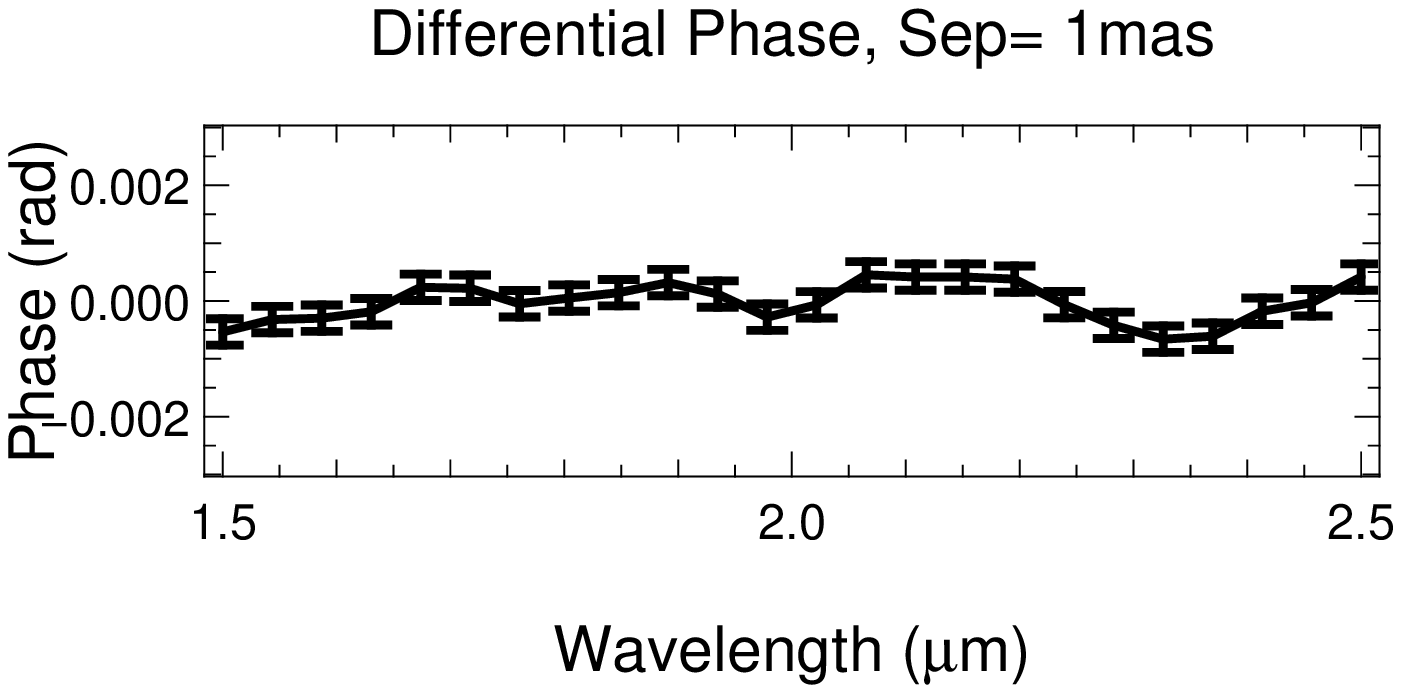}\\
    \includegraphics[width=0.31\textwidth, angle=0]{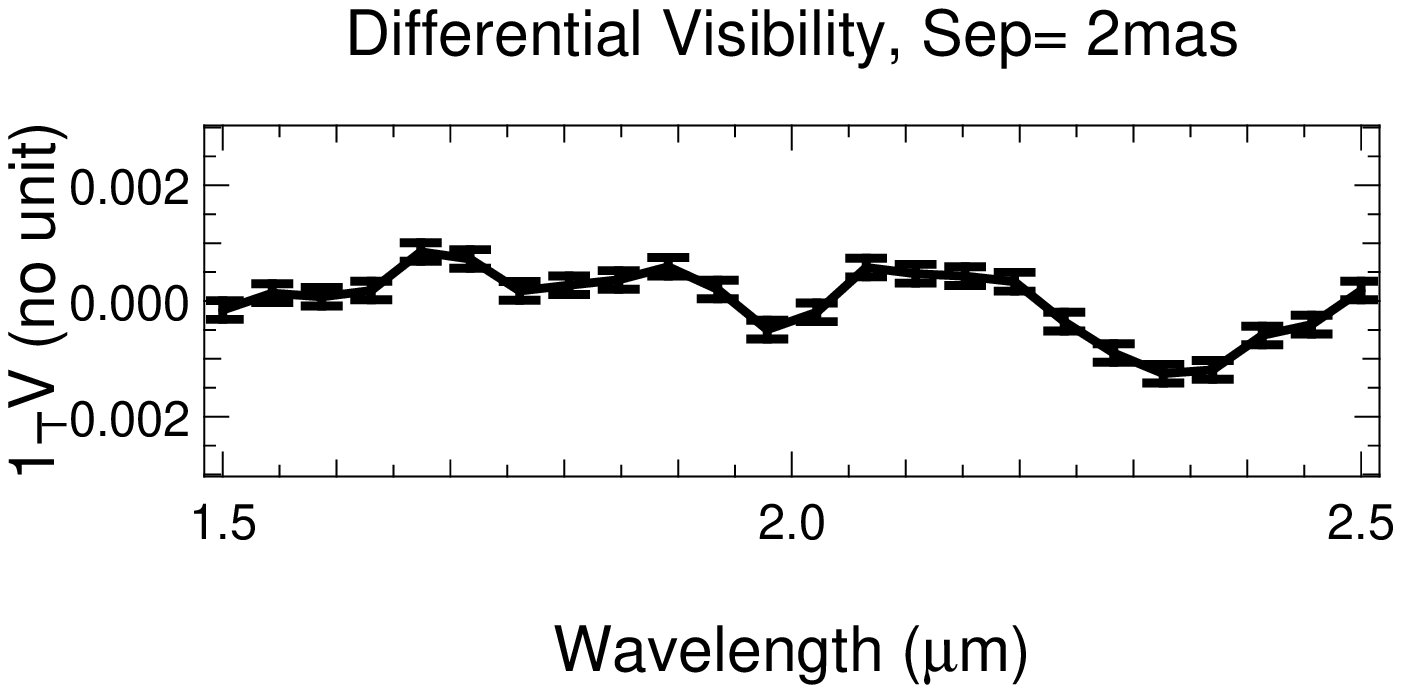}&
    \includegraphics[width=0.31\textwidth, angle=0]{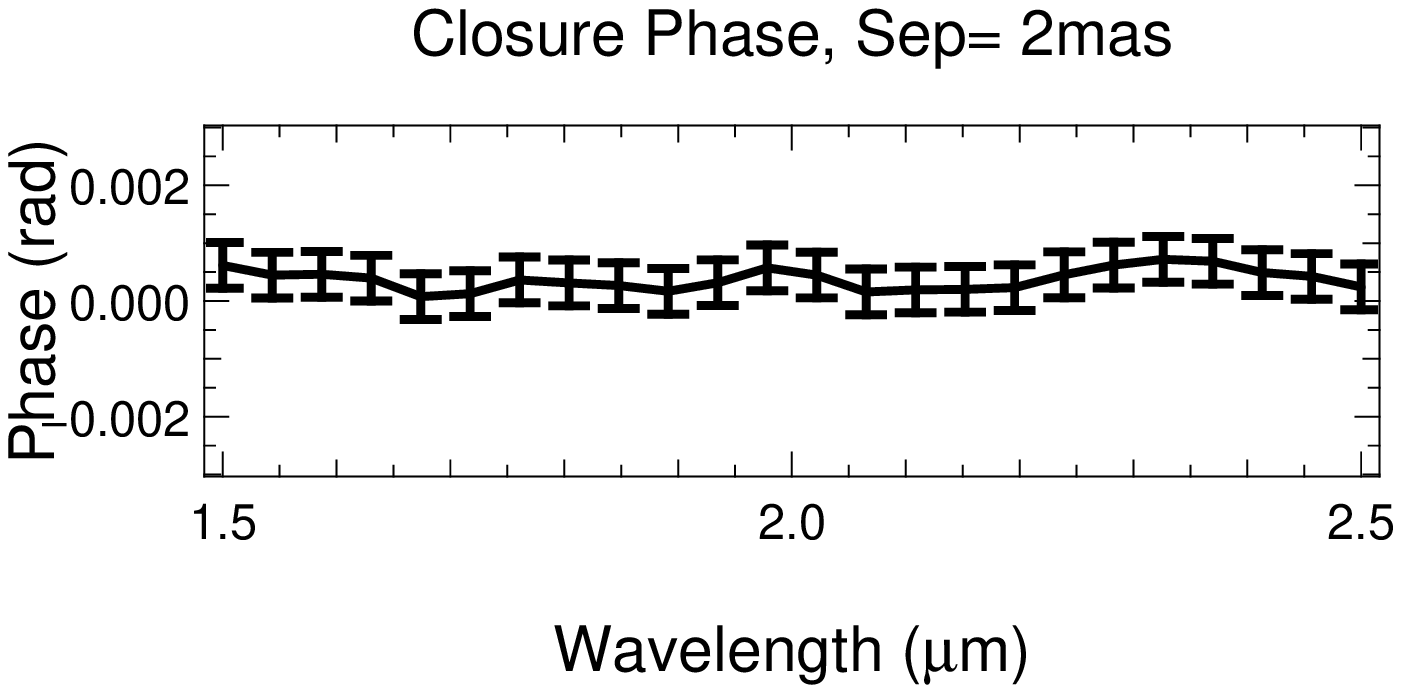}&
    \includegraphics[width=0.31\textwidth, angle=0]{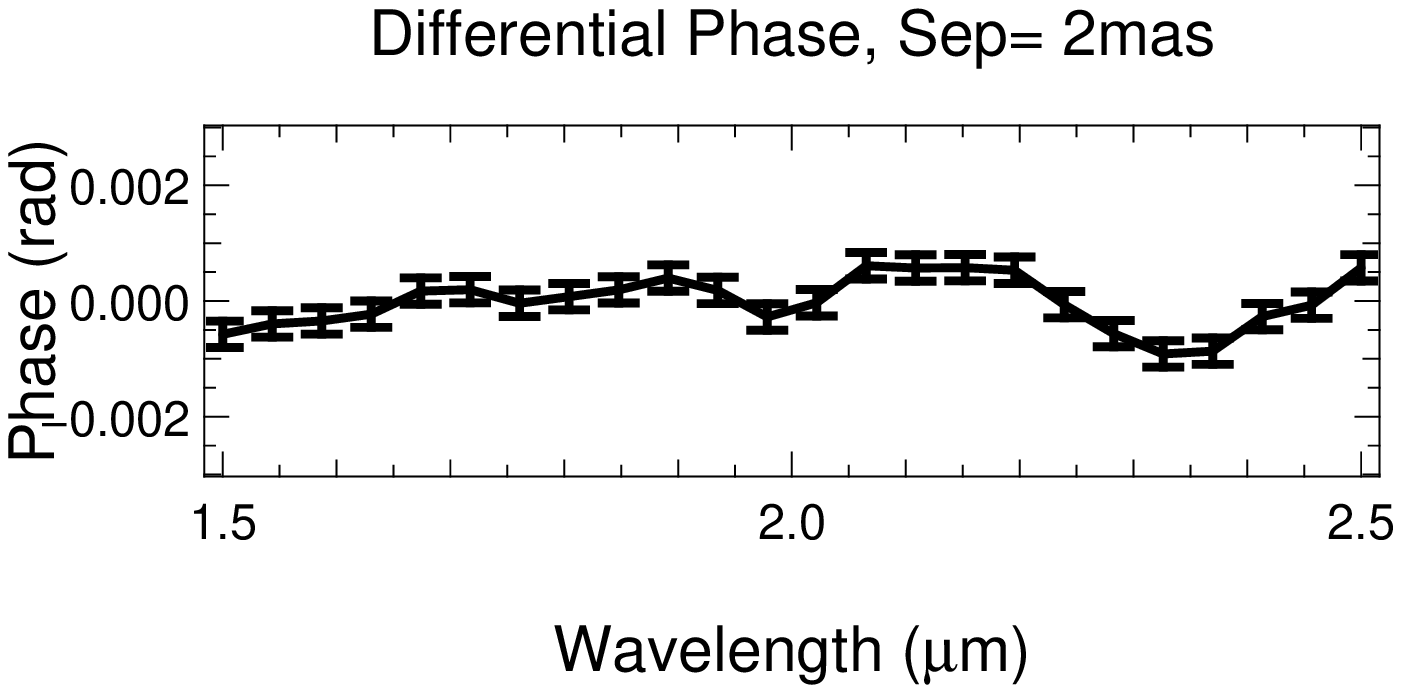}\\
    \includegraphics[width=0.31\textwidth, angle=0]{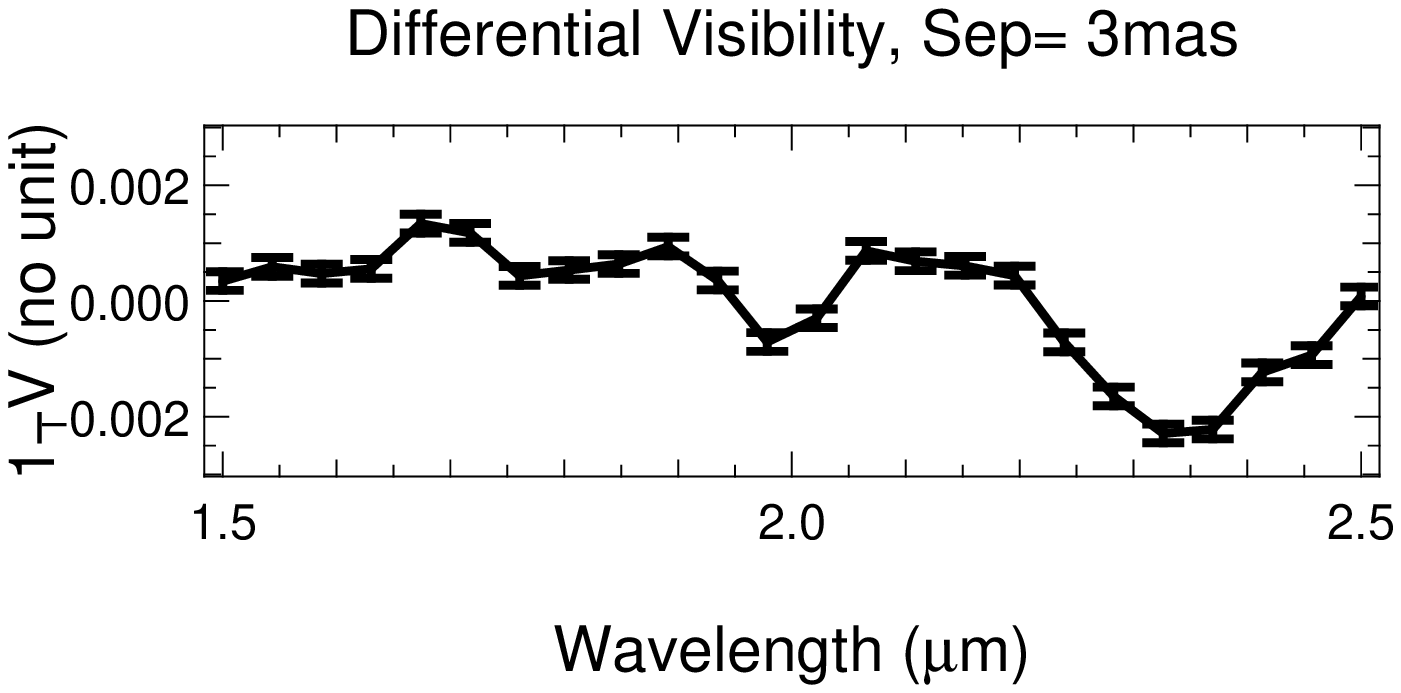}&
    \includegraphics[width=0.31\textwidth, angle=0]{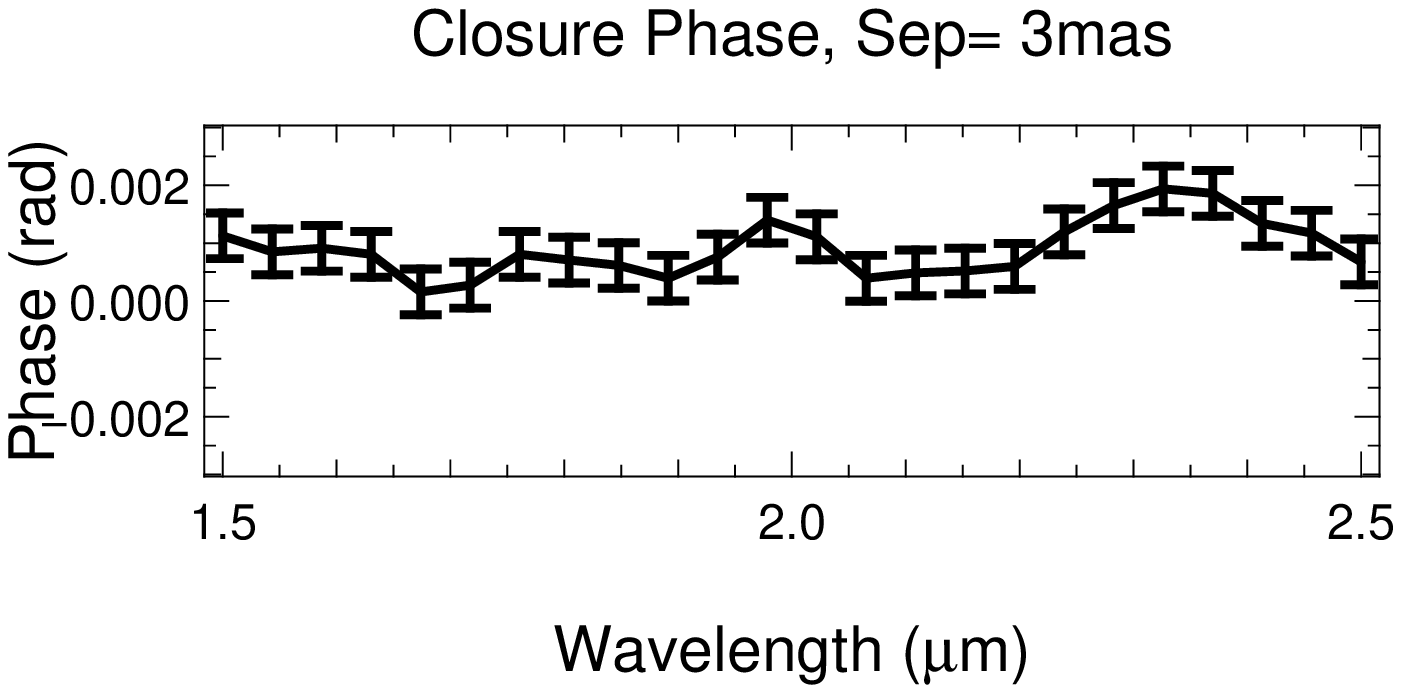}&
    \includegraphics[width=0.31\textwidth, angle=0]{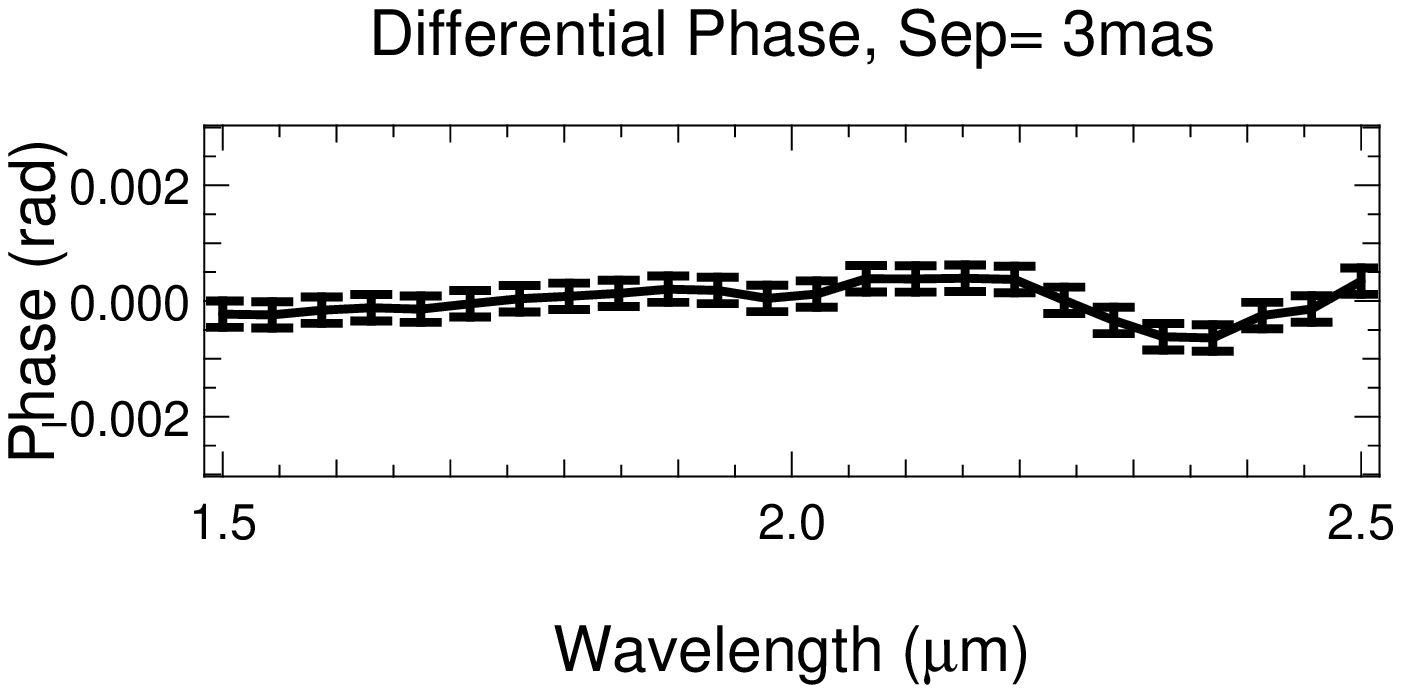}\\
  \end{tabular}
  \caption{
    \footnotesize{
      Simulations of an observation at VLTI
      with UT1-UT3-UT4 showing from left to right the differential
      visibility, the closure phase and the differential phase and
      from top to bottom separations ranging 1 to 3 mas. It shows a
      clear signal for all observables at 3mas but none for the
      closure phase at 1mas whereas the visibility and phase still have
      detectable signal. Photon amount is $4.2 \times 10^7$,
      detector noise $11 e^-$, number of frames 1000, number of pixels
      32 and average visibility 50\%. These plots evidence the
      ``super resolution'' properties of differental phase and
      visibility relatively to closure phase.
    }
  }
  \label{figSpectrum}
\end{figure}

\subsection{Results}

We tested the obtained signal for several Star/Companion separations
and found that the signal detection at $3 \sigma$ would occur for a
separation of only 1mas for differential phase and visibility and
2.5mas for closure phase. This effects is due to the fact that if we
perform a $1^{st}$ order taylor expansion of the phase, we get a
linear dependence with $\overrightarrow{u_{jk}} \cdot
\overrightarrow{\rho}$ whereas for visibility, we get a squared
dependence and for the closure phase we get a cubic dependence. So for
small separations, we have $\phi_{\rm diff} > V_{\rm diff} > \psi$.

\section{Best performances of AMBER}\label{sec:obs}

We used the AMBER / VLTI instrument to observe the bright calibration
star HD70060 at low spectral resolution during the GTO run
of 25 december 2004. We used the technique explained below to compute the
differential phases and differential visibilities and computed
statistical error bars. These ones were compared to the
theoretical ones assuming a detector noise of 11 electrons and a null
thermal noise. The figure \ref{figSpectrum} shows the resulting
average differential phases for 5 successive exposures where we
selected 50\% of the best frames using the finge SNR as selection
criterion. Each exposure represents about 20s of observation. They are
separated by about 60 s.

The standard deviation $\sigma$ in each exposure is the statistical dispersion of the
differential phase per spectral channel as described in
eq. \ref{statNoise}. We performed weighted averages using the SNR on
the fringe signal as weights. $n_\phi$ is the total number of
collected photons per spectral channel and $\sigma_\phi$ is the
accuracy expected from measured flux, detector noise and a supposed
object visibility of 1 (which means the measured visibility is
supposed to be only the instrumental one). 

% The difference between $\sigma$ and $\sigma_\phi$ is probably
% partially due to the large number of frames with very low contrast
% because of the VLTI/UT vibration (the laboratory instrumental contrast
% is measured at about 0.85 and the average ``instrumental'' contrast on
% the sky is 0.40). This gap is likely to be reduced in the future as
% vibrations are improved and/or a fringe tracker is used to stabilize
% the fringes. 

\begin{figure}[htbp]
  \centering
  \begin{tabular}{cc}
    \includegraphics[width=0.5\textwidth, angle=0]{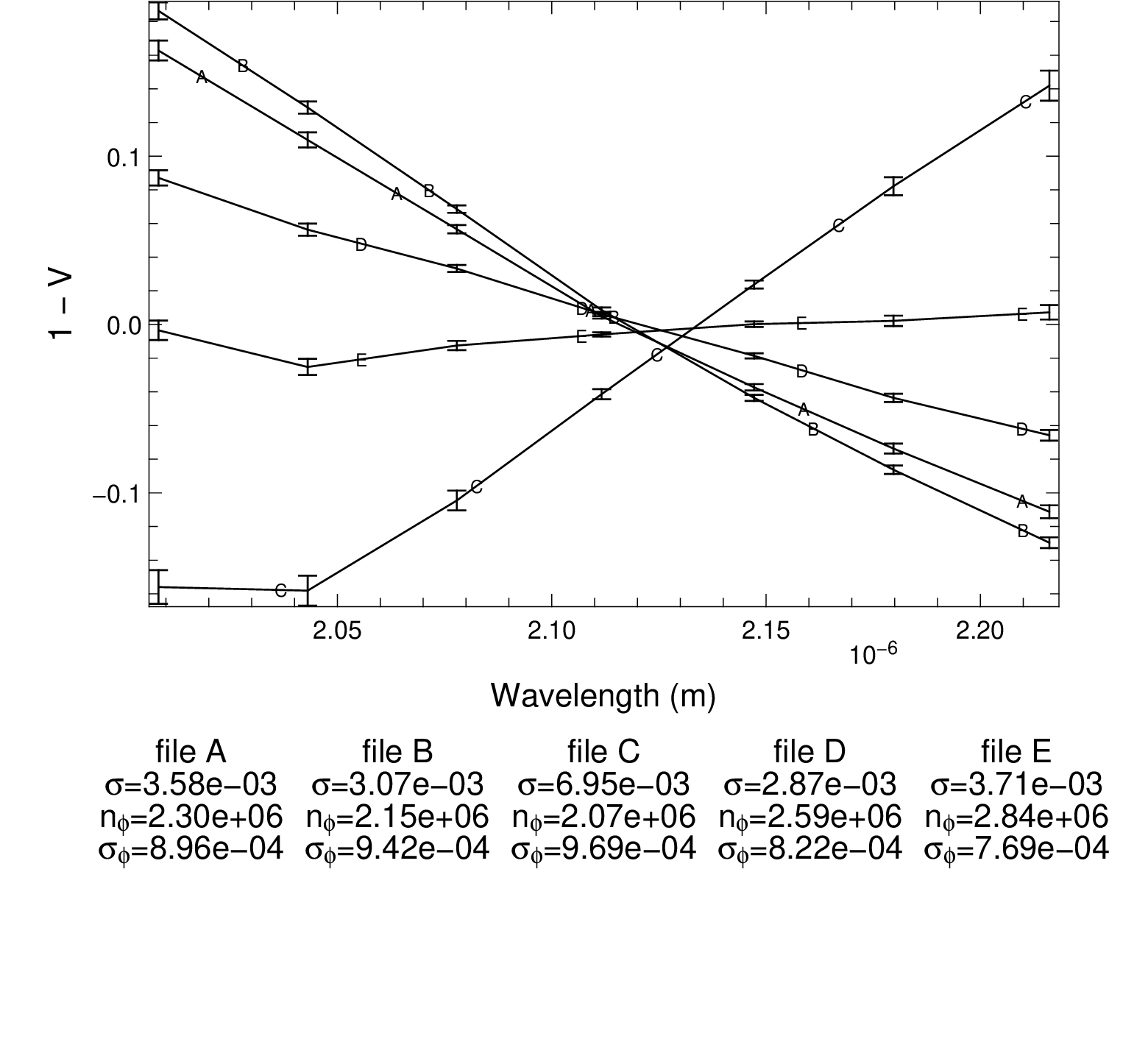}&
    \includegraphics[width=0.5\textwidth, angle=0]{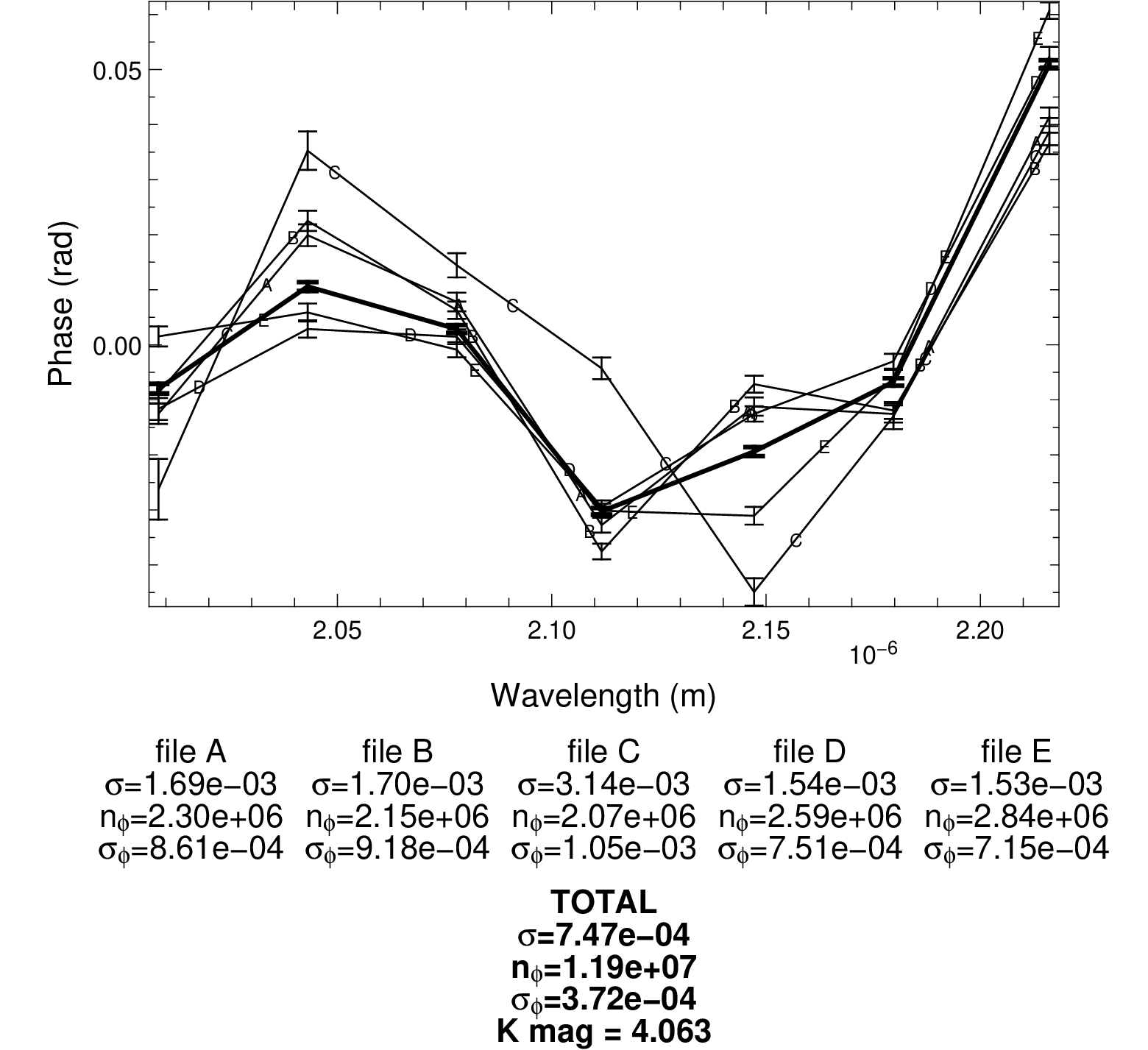}\\
  \end{tabular}
  \caption{
    \footnotesize{Differential Visibilities and Phases of the
      calibrator star HD70060 for 5 successive exposures (lines with letters) and
      the resulting average of the 5 exposures (thick line).
    }
  }
  \label{figSpectrum}
\end{figure}

\subsection{Differential Visibility}

The differential visibility values vary only by $4 \times 10^{-3}$
with time within each exposure but the general slope of the curve
changes dramatically between exposure, leading to a variation of about
0.05 radians rms over 5 minutes. This variation is dominated by the
changes in the achromatic piston jitter due to seeing and vibrations
fluctuations. In the current VLTI situation we have no tool to correct
this effect, except including an estimation of the exposure jitter in
the model fitting, with an impact on the SNR which cannot be
estimated. So the differential visibility is currently not usable for
very high accuracy applications. This situation will change
dramatically when a fringe tracker is operational. Then we will be
affected only by the residual piston jitter after fringe tracking
correction.

% The differential visibility seems to be highly dominated by
% differential jitter which affects the slope of the visibility versus
% wavelength. This is why we get a $10^{-3}$ radians error on each
% exposure and only $10^{-2}$ radians for all the exposures. This jitter
% effect could be corrected by a fringe tracker beforehand or by a
% frame-by-frame modelization and correction of the visibility loss due
% to the jitter in data processing. The use of Beam Commutation
% \citep{2005-ESP-MNRAS} is of no help in this case since it affects
% only the fringes phases and not the fringes amplitude. That is why in
% the current situation, even if we are at an individual exposure level
% of noise of about $10^{-3}$, we do not expect to improve this until a
% fringe tracker works on the VLTI and/or we find an efficient way of
% dynamically remove the jitter effect in the data processing.

\subsection{Differential Phase}

The rms variation of the measures within each exposure is of typically
1.8 milliradians (i.e. $\approx 1 \mu$arcsecond in colour-differential
astrometry). This is about two times the expected rms from fundamental
noise. Over the total 5 minutes we get 0.9 milliradians, again very
close to twice the photon noise. This shows that the different
measures seems statistically independent. An average of 1200 such
exposures (15 hours) is needed to reach the $0.5 \times 10^{-4}$
accuracy needed for the spectroscopy of $\tau$ Bootis b. The brighter
planet considered in \S \ref{sec:esp} could be observed in 1 hour.

 With the improvement of the VLTI (less vibrations, improved
 overheads) we could expect to use almost all frames instead of only
 50\% of them with an average instrumental contrast improved by a
 factor 2. Then the $\tau$ Boo observation would be achievable in a
 couple of hours. We also see a pattern as a function of lambda with a
 $10^{-2}$ radians rms over the K band. This pattern is stable over
 the 5 minutes of observations considered here. That means that it can
 be eliminated by a fast calibration cycle. We think that it is a
 mixture of atmospheric dispersion and measurement
 effects. Measurement effects can be eliminated by beam
 commutation. Atmospheric dispersion will be eliminated in the closure
 phase and we plan to try to fit it in the differential phase. 

% a systematic pattern appears on the differential phase. It explains
% the global $10^{-3}$ radians in 5 minutes. The pattern might be due to
% differential dispersion, but the closure phase for this particular
% exposure is too noisy ($\approx 10^{-2}$ radians) to conclude. After 5
% minutes and at the $10^{-3}$ radians level, the differential phases
% seems to be statistically independent.

% The current situation can be
% summarized as follows: With the current slow calibration using an
% external reference star, we are roughly limited to $10^{-3}$ radians
% accuracy at a $1 \sigma$ level, sufficient for the study of, for
% example, circumstellar material. If extrapolated to a few hours of
% observations with a speeded up (20') science-calibrator cycle, the
% process of careful exposure selection, drift fit and external
% calibration, would probably yield $10^{-3}$ radians at several
% $\sigma$ level. If we compare these results with the precedent
% simulations, this would be sufficient for a few brown dwarf
% candidates or very hot Extra Solar Giant Planets. If it appears to
% reach the expected performances, the Beam Commuting Device
% \citep{2005-ESP-MNRAS} should allow to make the individual 30 seconds
% exposures completely statistically independent. Then in a few hundreds
% of minutes, we would reach the $3\sigma < 10^{-4}$ radians precision,
% sufficient for the spectroscopy of Pegasides.

\subsection{Closure Phase}

In this relatively poor quality early data, we have too little frames
where the three fringe patterns are good enough. This explains why the
closure phase is much more noisy (typically $10^{-2}$ radians rms) than the
differential phases. We are therefore unable to say what part of the
differential phase pattern is due to differential chromatic OPD.

\section{Conclusion}

% We showed in this article that without specific instrumental
% calibration, we can reach quite easily a $10^{-3}$ radians precision
% on the differential phase. Moreover, different exposures separated in
% time seems to be statistically independent and allow a direct
% extrapolation to longer integration times neglecting all the
% instrumental drifts. It show that we could reach in a reasonable
% amount of time ($\approx 20h$) precisions as high as $10^{-4}$ radians
% the use of beam commutation would possibly allow to calibrate all the
% instrumental drifts effects that would appear for longer exposure,
% opening the field of Giant Hot Extrasolar Planets spectroscopy. We
% showed also in this article that with the current performances of the
% AMBER instrument, we can reach quite easily a $10^{-3}$ radians
% precision on the differential phase, leading to the possibility to
% detect faint companions very close to its parent star (with a
% separation as close as 1mas). 

The preliminary data reduction of bright sources observed in low
spectral resolution with AMBER shows that the measured differential
phases are accurate and stable enough to achieve the spectroscopy and
angular separation of the most favorable Pegasi planets in a few 15
hours observations. This value should be reduced to 2 hours with the
foreseen simple improvements of the VLTI. The resulting spectra would
be affected by an instrumental term and/or an atmospheric chromatic
differential OPD term producing a smooth $10^{-2}$ radians pattern over the
K band.

 When the instrumental term will be eliminated by beam
commutation, the remaining differential OPD might be possible to fit
in the data reduction procedure. However, only a successful use of
closure phase guarantees the elimination of the differential OPD. The
current quality of the VLTI does not allow accurate closure phase
measurements, but this should be improved soon, when the three fringe
pattern are better stabilized. We remain very optimistic about the
possibility to do spectroscopy of Pegasi planets with AMBER quite
sson.

\begin{acknowledgments}
  The data presented here was taken at the Paranal Observatory in
  Chile within the AMBER Guaranteed Time.

 We thank all the consortium
  members listed in {\ttfamily http://amber.obs.ujf-grenoble.fr}.
\end{acknowledgments}

\bibliography{biblio}%>>>> bibliography data in report.bib
\bibliographystyle{aa}%>>>> makes bibtex use spiebib.bst

\end{document}